\documentclass{aa}
\usepackage{graphicx}
\usepackage{txfonts}
\defcitealias{dep19a}{Paper~I}
\defcitealias{dep19b}{Paper~II}
\begin{document}

\title{Plasma-environment effects on K lines of astrophysical interest}

\subtitle{III. IPs, K thresholds, radiative rates, and Auger widths in \ion{Fe}{ix} -- \ion{Fe}{xvi}}

\author{J. Deprince
          \inst{1},
          M.~A. Bautista
           \inst{2},
          S. Fritzsche
          \inst{3,4},
          J.~A.~Garc\'ia
          \inst{5,6},
          T.~R. Kallman
          \inst{7},\\
          C. Mendoza
          \inst{2},
          P. Palmeri
          \inst{1}
          \and
          P. Quinet\inst{1,8}
}

\institute{Physique Atomique et Astrophysique, Universit\'e de Mons -- UMONS, B-7000 Mons, Belgium\\
              \email{patrick.palmeri@umons.ac.be}
         \and
             Department of Physics, Western Michigan University, Kalamazoo, MI 49008, USA
         \and
             Helmholtz Institut Jena, 07743 Jena, Germany
         \and
             Theoretisch Physikalisches Institut, Friedrich Schiller Universit\"at Jena,
             07743 Jena, Germany
         \and
             Cahill Center for Astronomy and Astrophysics, California Institute of Technology, Pasadena, CA 91125, USA
         \and
             Dr. Karl Remeis-Observatory and Erlangen Centre for Astroparticle Physics, Sternwartstr.~7, 96049 Bamberg, Germany
         \and
             NASA Goddard Space Flight Center, Code 662, Greenbelt, MD 20771, USA
         \and
             IPNAS, Universit\'e de Li\`ege, Sart Tilman, B-4000 Li\`ege, Belgium
}
\titlerunning{Plasma effects on K lines in \ion{Fe}{xvi} -- \ion{Fe}{ix}}

\authorrunning{Deprince et al.}

\date{Received ??; accepted ??}

\abstract
   {}
   {In the context of black-hole accretion disks, we aim to compute the plasma-environment effects on the atomic parameters used to model the decay of K-vacancy states in moderately charged iron ions, namely \ion{Fe}{ix} -- \ion{Fe}{xvi}.}
   {We used the fully relativistic multiconfiguration Dirac--Fock (MCDF) method approximating the plasma electron--nucleus and electron--electron screenings with a time-averaged Debye-H\"uckel potential.}
   {We report modified ionization potentials, K-threshold energies, wavelengths, radiative emission rates, and Auger widths for plasmas characterized by electron temperatures and densities in the ranges $10^5{-}10^7$~K and $10^{18}{-}10^{22}$~cm$^{-3}$.}
   {This study confirms that the high-resolution X-ray spectrometers onboard the future {\it XRISM} and {\it ATHENA} space missions will be capable of detecting the lowering of the K edges of these ions due to the extreme plasma conditions occurring in accretion disks around compact objects.}


\maketitle
%

\section{Introduction}

The present paper is the third in a series devoted to density effects on atomic parameters used to model K-vacancy
states in ions of astrophysical interest; namely, ionization potentials, K thresholds, transition wavelengths, radiative emission rates, and Auger widths. \citet[][hereafter Paper~I]{dep19a} computed such data for the oxygen isonuclear sequence and \citet[][hereafter Paper~II]{dep19b} did so for highly ionized Fe ions (\ion{Fe}{xvii} -- \ion{Fe}{xxv}) using the relativistic multiconfiguration Dirac--Fock (MCDF) method \citep{gra80,mck80,gra88}, as implemented in the GRASP92 \citep{par96} and RATIP \citep{fri12} atomic structure packages. The plasma electron--nucleus and electron--electron shielding were approximated with a time-averaged Debye-H\"uckel (DH) potential. Here we report our calculations using this method for the third-row Fe ions, \ion{Fe}{ix} -- \ion{Fe}{xvi}.

The absorption and emission of high-energy photons occur in the dense plasma ($10^{21}{-}10^{22}$~cm$^{-3}$) of the inner region of the accretion disks associated with compact objects, such as black holes and neutron stars. The modeling of the associated X-ray spectra, which can be observed with space telescopes such as {\it Chandra}, {\it XMM-Newton}, {\it Suzaku}, and {\it NuSTAR}, gives a measure of the composition, temperature, and degree of ionization for the plasma \citep{ros05, gar10}. For a black hole, for instance, the distortion of the Fe K lines by the strong relativistic effects constrains its angular momentum  \citep{rey13, gar14}. However, most of the atomic data used in the modeling of K-shell processes do not take density effects into account, compromising their significance for abundance determinations beyond densities of $10^{18}$~cm$^{-3}$ \citep{smi14}.

\section{Theoretical approach}

The MCDF formalism was already described in \citetalias{dep19a} and \citetalias{dep19b}, so in this paper, we only aim to provide brief indications of the DH modifications that were introduced to handle weakly coupled plasmas. The DH screened Dirac--Coulomb Hamiltonian \citep{sah06} takes the form of
\begin{equation}
H^{DH}_{DC}=\sum_i c \vec{\alpha_i} \cdot \vec{p_i}+ \beta_i c^2 - \frac{Z}{r_i}
e^{-\mu r_i}
+ \sum_{i>j} \frac{1}{r_{ij}} e^{-\mu r_{ij}}\ ,
\label{dh}
\end{equation}
where $r_{ij}=|\vec{r}_i-\vec{r}_j|$ and the plasma screening parameter $\mu$ is the inverse of the Debye shielding length $\lambda_D$, which can be expressed in atomic units (au) as a function of the plasma electron density $n_e$ and temperature $T_e$ as
\begin{equation}
\mu = \frac{1}{\lambda_D} = \sqrt{\frac{4\pi n_e}{k T_e}}\ .
\label{screen}
\end{equation}
Typical plasma conditions in black-hole accretion disks are $T_e\sim  10^5{-}10^7$~K and $n_e\sim 10^{18}{-}10^{22}$~cm$^{-3}$ \citep{sch13} which, for weakly coupled plasmas correspond to screening parameters of $0.0\leq \mu\leq 0.24$~au, and, for a completely ionized hydrogen plasma (plasma ionization $Z^*=1$) correspond to plasma coupling parameters of
\begin{equation}
\Gamma = \frac{e^2}{4\pi\epsilon_0 d kT_e}
,\end{equation}
with
\begin{equation}
d=\left(\frac{3}{4\pi n_e}\right)^{1/3}
,\end{equation}
in the range $0.0003\leq \Gamma \leq 0.3$.

Following \citetalias{dep19b}, the MCDF expansions for \ion{Fe}{ix} -- \ion{Fe}{xvi} are generated using the active space  method, whereby electrons from the reference configurations in Table~\ref{as} are singly and doubly excited to configurations that include $n=3$ and ${\rm 4s}$ orbitals. For instance, the non-relativistic configuration ${\rm [1s2s]3p^63d4s}$, generated by a double electron excitation from the 1s and 2s active subshells of the reference configuration ${\rm 3p^6}$ to the 3d and 4s active orbitals, is included in the expansions of the atomic state functions (ASFs) in \ion{Fe}{ix}. As defined in Eq.~(1) of  \citetalias{dep19b}, the number of configuration state functions (CSFs) generated for the MCDF expansions of the ASFs  are also given in Table~\ref{as} for each ion. We consider plasma screening parameters in the range $0.0\leq\mu\leq 0.25$~au, the upper-limit choice corresponding to the extreme plasma conditions found in accretion disks.

\begin{table*}[!ht]
  \caption{Reference configurations and active orbital sets used to build up the MCDF active space by single and double electron excitations to the corresponding active orbital sets along with the number of configuration state functions (CSFs) generated for the MCDF expansions in \ion{Fe}{ix} -- \ion{Fe}{xvi}. \label{as}}
  \centering
  \small
  \begin{tabular}{lllr}
  \hline\hline
  \noalign{\smallskip}
  Ion & Reference configurations & Active orbital set & Number of CSFs\\
  \hline
  \noalign{\smallskip}
                \ion{Fe}{ix} & ${\rm 3p^6}$,
                ${\rm [3p]3d}$,
                ${\rm [2p]3d}$,
                ${\rm [1s]3d}$
                & ${\rm \{1s,2s,2p,3s,3p,3d,4s\}}$
                & 20009  \\
                \ion{Fe}{x} & ${\rm 3p^5}$,
                ${\rm [2p]3p^6}$,
                ${\rm [1s]3p^6}$                
                & ${\rm \{1s,2s,2p,3s,3p,3d,4s\}}$
                & 6312  \\
                \ion{Fe}{xi} & ${\rm 3p^4}$,
                ${\rm [2p]3p^5}$,
                ${\rm [1s]3p^5}$
            & ${\rm \{1s,2s,2p,3s,3p,3d,4s\}}$
                & 12981  \\
                \ion{Fe}{xii} & ${\rm 3p^3}$,
            ${\rm [2p]3p^4}$,
            ${\rm [1s]3p^4}$
            & ${\rm \{1s,2s,2p,3s,3p,3d,4s\}}$
                & 37967  \\
                \ion{Fe}{xiii} & ${\rm 3p^2}$,
                ${\rm [2p]3p^3}$,
                ${\rm [1s]3p^3}$
                & ${\rm \{1s,2s,2p,3s,3p,3d,4s\}}$
                & 46771  \\
                \ion{Fe}{xiv} & ${\rm 3p}$,
                ${\rm [2p]3p^2}$,
                ${\rm [1s]3p^2}$
            & ${\rm \{1s,2s,2p,3s,3p,3d,4s\}}$
                & 35109  \\
                \ion{Fe}{xv} & ${\rm 3s^2}$,
                ${\rm [2p]3s^23p}$,
                ${\rm [1s]3s^23p}$
            & ${\rm \{1s,2s,2p,3s,3p,3d,4s\}}$
                & 16853  \\
                \ion{Fe}{xvi} & ${\rm 3s}$,
                ${\rm [2p]3s^2}$,
                ${\rm [2p]3s3p}$,
                ${\rm [1s]3s^2}$,
                ${\rm [1s]3s3p}$
                & ${\rm \{1s,2s,2p,3s,3p,3d,4s\}}$
                & 25914  \\
  \hline
  \end{tabular}
\end{table*}

\section{Results and discussion}

\subsection{Ionization potentials and K thresholds}

The computed ionization potentials (IPs) and K thresholds ($E_K$)  are given in Table~\ref{IP-plasma} and Table~\ref{Kth-plasma}, respectively, for plasma screening parameter $\mu=0.0$, 0.1 and 0.25. The case $\mu=0.1$ corresponds to plasma conditions of $T_e = 10^5$~K and $n_e = 10^{21}$~cm$^{-3}$ and $\mu = 0.25$ au to $T_e = 10^5$~K and $n_e = 10^{22}$ cm$^{-3}$. For the isolated ion case ($\mu = 0.0$ au), the computed IPs are compared in Table~\ref{IP-plasma} with the values quoted in the atomic database \citep{nist} at the National Institute of Standards and Technology (NIST), showing an agreement within 0.1\%.

\begin{table}[!ht]
  \caption{Computed ionization potentials for \ion{Fe}{ix} -- \ion{Fe}{xvi} as a function of the plasma screening parameter $\mu$ (au). NIST values are also listed for comparison. \label{IP-plasma}}
  \centering
  \small
  \begin{tabular}{llccc}
  \hline\hline
  \noalign{\smallskip}
  Ion    & \multicolumn{4}{c}{IP (eV)} \\
  \cline{2-5}
  \noalign{\smallskip}
   &  NIST$^a$ & $\mu = 0.0$ & $\mu = 0.1$ & $\mu = 0.25$ \\
  \hline
  \noalign{\smallskip}
  \ion{Fe}{ix}   & $233.6(4)$    & 230.91 &     206.99 & 173.48 \\
  \ion{Fe}{x}    & $262.10(12)$  & 263.14 &     236.65 & 199.74 \\
  \ion{Fe}{xi}   & $290.9(4)$    & 294.15 & 265.04 & 224.58 \\
  \ion{Fe}{xii}  & $330.8(6)$    & 325.81 & 294.04 & 249.75 \\
  \ion{Fe}{xiii} & $361.0(7)$    & 356.95 & 322.54 & 274.48 \\
  \ion{Fe}{xiv}  & $392.2(7)$    & 388.72 & 351.68 & 300.03 \\
  \ion{Fe}{xv}   & $456.2(5)$    & 457.14 & 417.57 & 362.77 \\
  \ion{Fe}{xvi}  & $489.312(14)$ & 488.86 & 446.68 & 388.34 \\
  \hline
  \end{tabular}
   \tablefoot{\tablefoottext{a}{\citet{nist}.}}
\end{table}

\begin{table}[!ht]
  \caption{Computed K-thresholds for \ion{Fe}{ix} -- \ion{Fe}{xvi} as a function of the plasma screening parameter $\mu$ (au). \label{Kth-plasma}}
  \centering
  \small
  \begin{tabular}{lccc}
  \hline\hline
  \noalign{\smallskip}
  Ion & \multicolumn{3}{c}{$E_K$~(eV)} \\
  \cline{2-4}
  \noalign{\smallskip}
    & $\mu = 0.0$       & $\mu = 0.1 $ & $\mu = 0.25$ \\
  \hline
  \noalign{\smallskip}
  \ion{Fe}{ix}   & 7308.25 & 7283.14 & 7243.32 \\
  \ion{Fe}{x}    & 7351.72 & 7324.04 & 7280.88 \\
  \ion{Fe}{xi}   & 7393.76 & 7363.48 & 7316.89 \\
  \ion{Fe}{xii}  & 7434.46 & 7401.52 & 7351.12 \\
  \ion{Fe}{xiii} & 7483.90 & 7448.32 & 7394.27 \\
  \ion{Fe}{xiv}  & 7535.85 & 7497.64 & 7439.95 \\
  \ion{Fe}{xv}   & 7591.60 & 7550.76 & 7489.46 \\
  \ion{Fe}{xvi}  & 7639.54 & 7596.05 & 7531.05 \\
  \hline
  \end{tabular}
\end{table}

\begin{figure}[!ht]
  \centering
  \includegraphics*[pagebox=mediabox, width=\columnwidth]{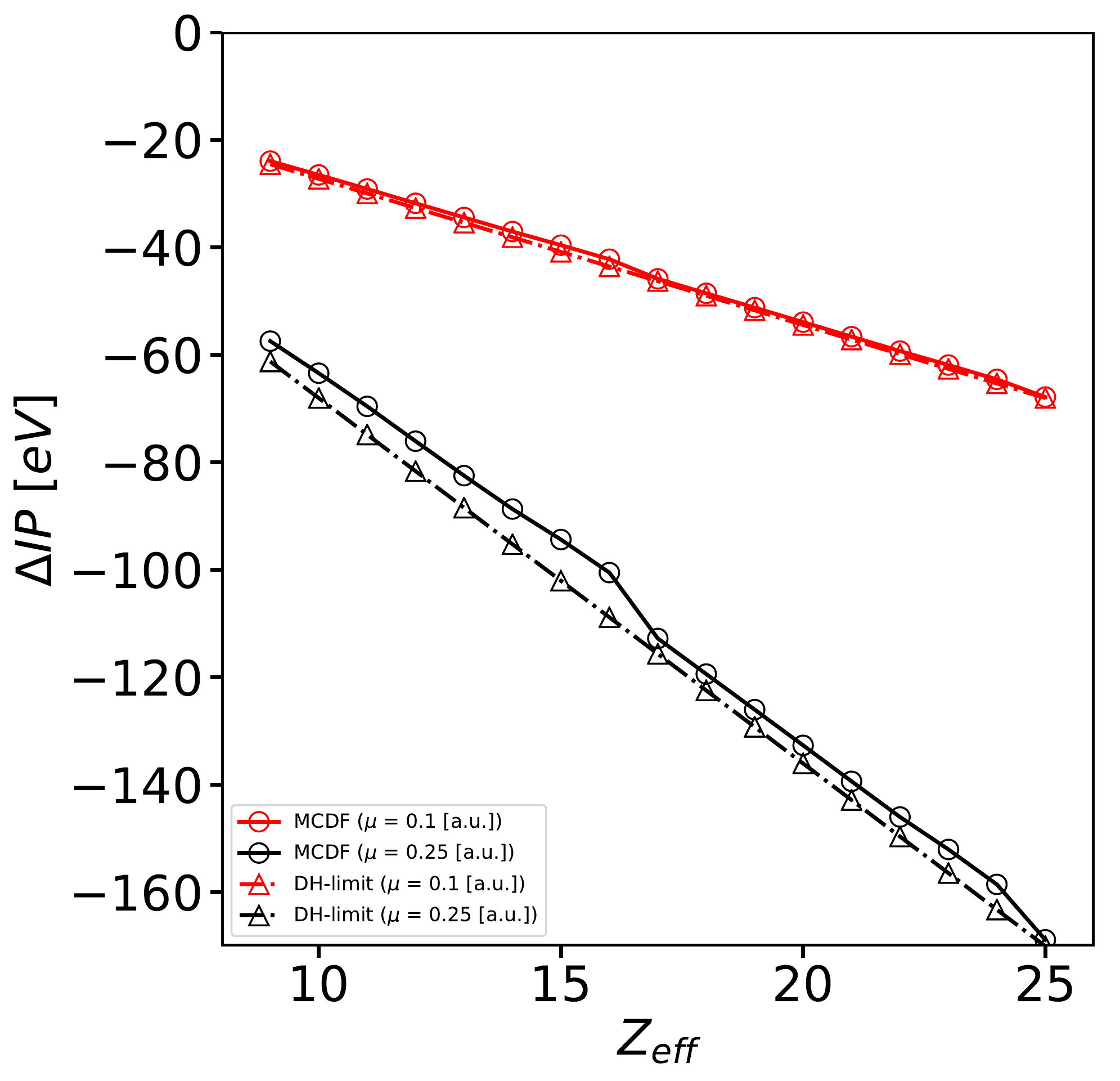}
  \caption{Ionization potential shifts, $\Delta IP$, in \ion{Fe}{ix} -- \ion{Fe}{xxv} as a function of the effective charge $Z_{\rm eff}=Z-N+1$. Red open circles: $\mu=0.1~{\rm au}$. Black open circles: $\mu=0.25~{\rm au}$. Open triangles: Debye-H\"uckel limit $\Delta {IP}_{\rm DH}=-Z_{\rm eff}\,\mu$.} \label{Fig1}
\end{figure}

\begin{figure}[!ht]
  \centering
  \includegraphics*[pagebox=mediabox, width=\columnwidth]{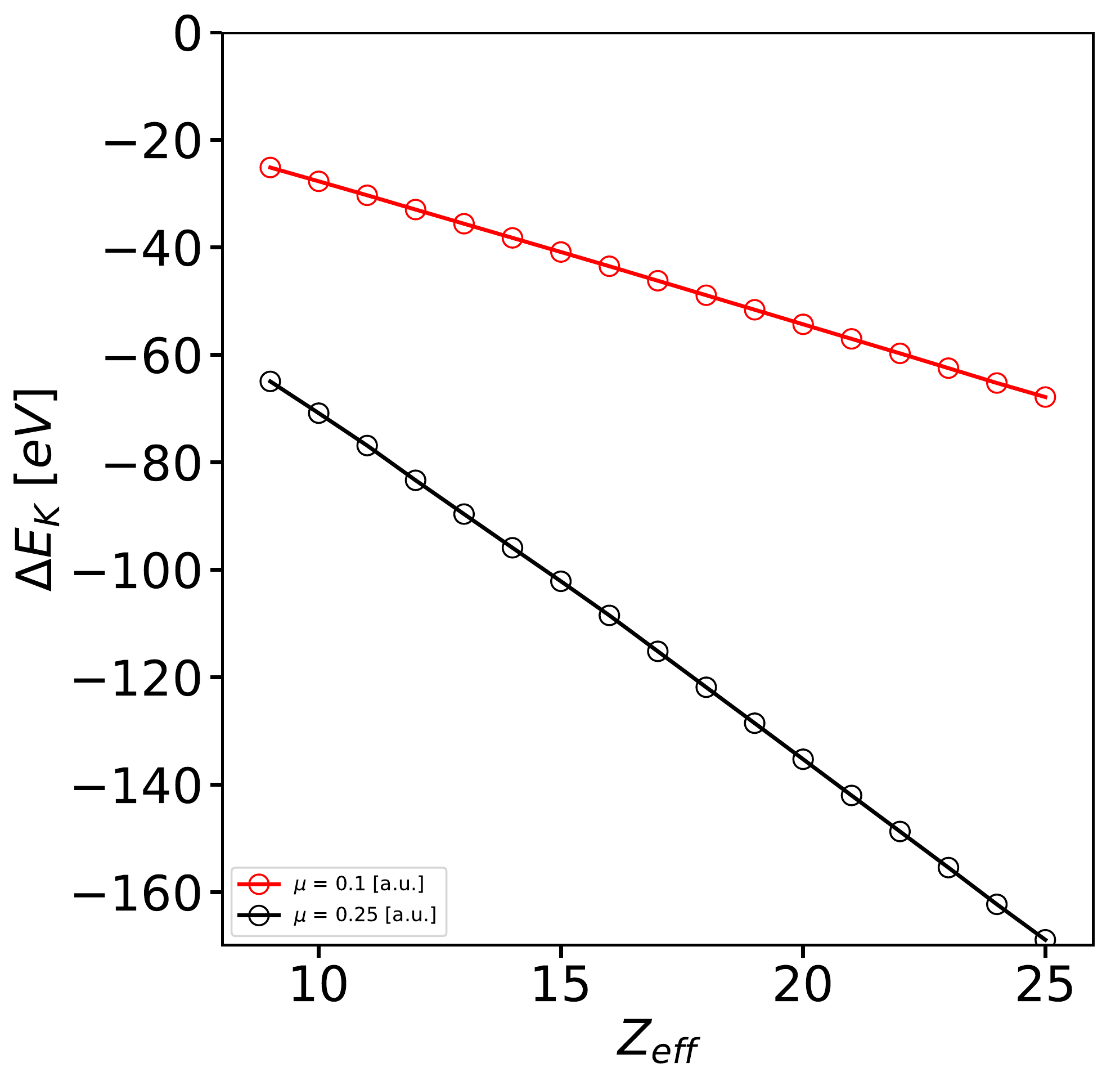}
  \caption{K-threshold shifts, $\Delta E_K$, in \ion{Fe}{ix} -- \ion{Fe}{xxv} as a function of the effective charge $Z_{\rm eff}=Z-N+1$. Red open circles: $\mu=0.1~{\rm au}$. Black open circles: $\mu=0.25~{\rm au}$.} \label{Fig2}
\end{figure}

Tables~\ref{IP-plasma}--\ref{Kth-plasma} also show the IP and K-threshold lowering by the plasma environment at $\mu = 0.1$ au and $\mu = 0.25$ au: the IP downshifts are, respectively, 9--10\% and 21--25\%, while those for the K thresholds, due to their much larger energy, only 0.4--0.6\% and 1\%. Moreover, in agreement with \citetalias{dep19b}, the absolute IP and K-threshold downshifts for each species are practically similar in magnitude. This effect can be further appreciated in Figs.~\ref{Fig1}--\ref{Fig2}, where we plot the absolute downshifts as a function of the effective charge, $Z_{\rm eff}=Z-N+1$ where $Z$ and $N$ are respectively the atomic and electron numbers of the ionic system. We also include in these figures the Fe species with $25\leq Z_{\rm eff}\leq 17$ from \citetalias{dep19b} and the Debye-H\"uckel limit $\Delta IP_{\rm DH}=-Z_{\rm eff}\,\mu$ \citep{sp66,crow14}.

\begin{figure}[!ht]
  \centering
  \includegraphics*[pagebox=mediabox, width=\columnwidth]{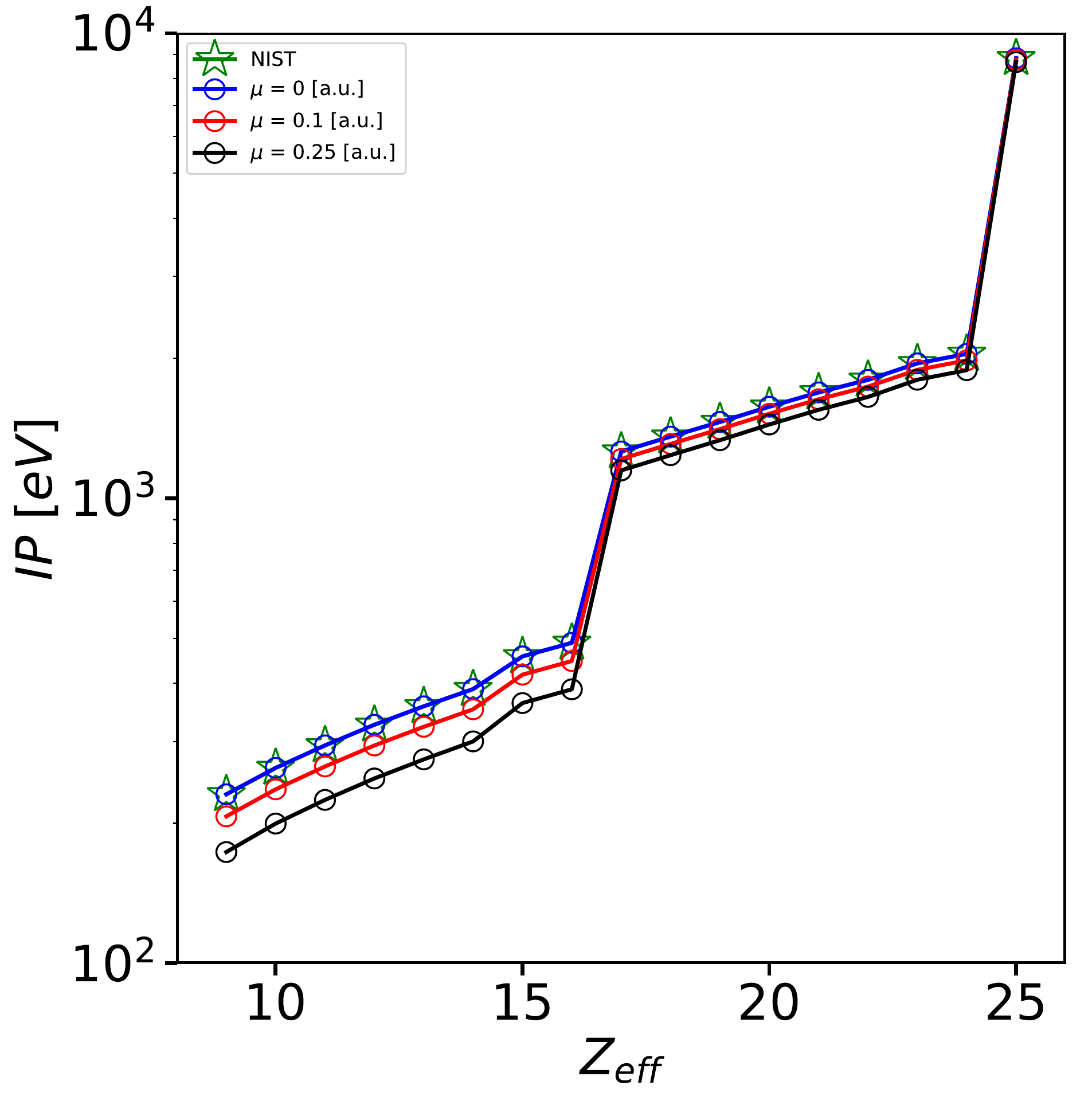}
  \caption{Ionization potential, IP, in \ion{Fe}{ix} -- \ion{Fe}{xxv} as a function of the effective charge $Z_{\rm eff}=Z-N+1$. Blue open circles: $\mu=0~{\rm au}$ (isolated atom case). Red open circles: $\mu=0.1~{\rm a.u}$. Black open circles: $\mu=0.25~{\rm au}$. Green open stars: NIST \citep{nist}.} \label{Fig3}
\end{figure}

\begin{figure}[!ht]
  \centering
  \includegraphics*[pagebox=mediabox, width=\columnwidth]{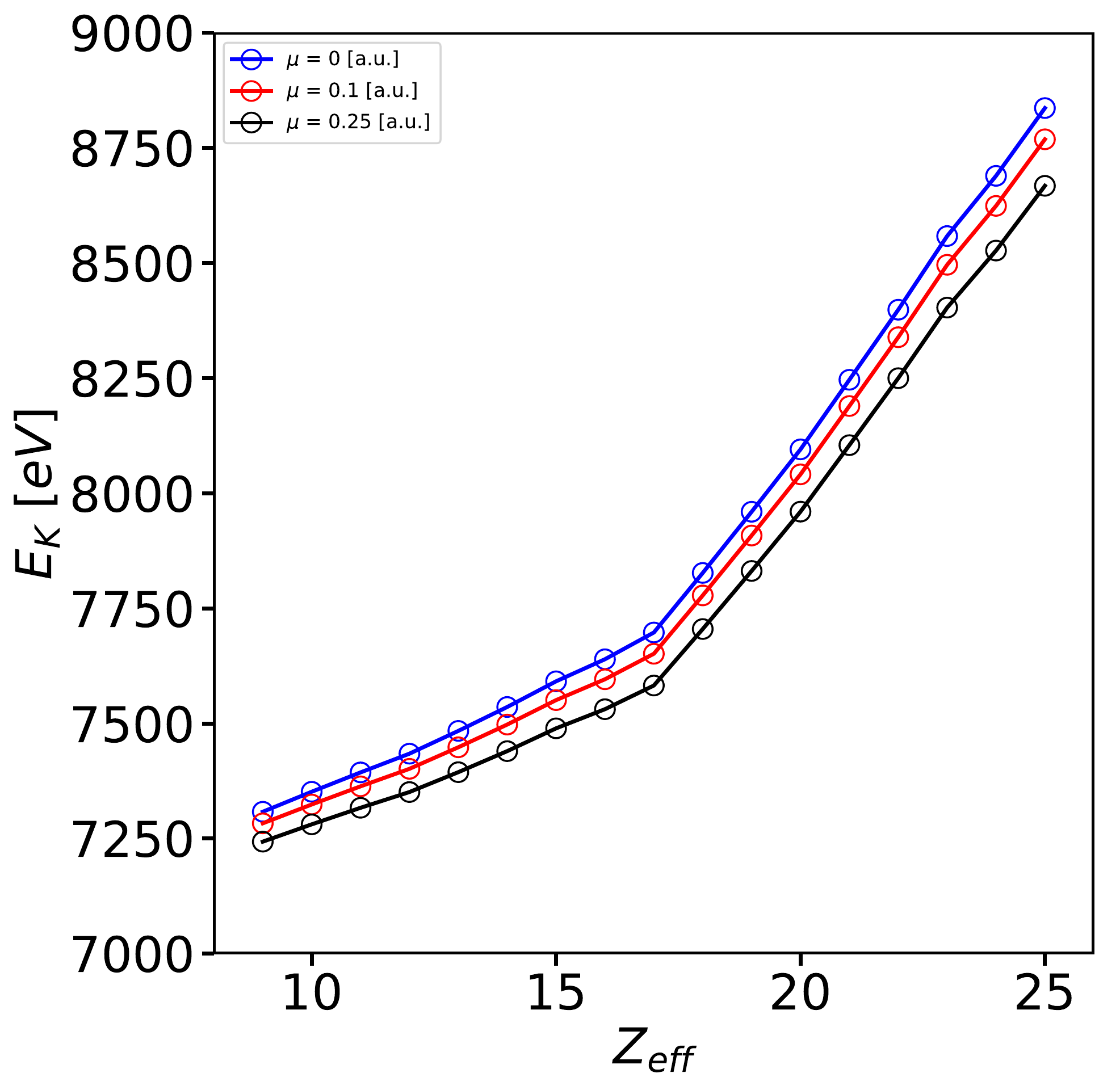}
  \caption{K thresholds, $E_K$, in \ion{Fe}{ix} -- \ion{Fe}{xxv} as a function of the effective charge $Z_{\rm eff}=Z-N+1$. Blue open circles: $\mu=0~{\rm au}$ (isolated atom case). Red open circles: $\mu=0.1~{\rm au}$. Black open circles: $\mu=0.25~{\rm au}$.} \label{Fig4}
\end{figure}

The linear lowerings of both the IP and K threshold, $\Delta IP$ and $\Delta E_K$ , respectively, with $Z_{\rm eff}$ and their close magnitude for each ion are hereby reiterated. We also verify that the Debye-H\"uckel limit is a good approximation of the IP lowering except for two discontinuities at $Z_{\rm eff}=17$ and 25 conspicuous at the higher plasma screening parameter ($\mu = 0.25$), which are caused respectively by the closing of the L and K shells. It can be seen in Fig.~\ref{Fig3} that the IP increases linearly with $Z_{\rm eff}$ but two large escalations, namely a factor of 2.6 and 4, occur respectively for the closed L- and K-shell species \ion{Fe}{xvii} and \ion{Fe}{xxv}. The behavior of the K threshold with effective charge is somewhat different (see Fig.~\ref{Fig4}); although it still increases linearly, the gradient becomes steeper at $Z_{\rm eff}=17$ and no effect is appreciable at  $Z_{\rm eff}=25$ since the K-shell electron is located deeper close to the nucleus in contrast to the relatively weakly bound valence electron.

\subsection{Radiative transitions}

The present MCDF K-line wavelengths are in excellent agreement with those obtained with the pseudo-relativistic Hartree--Fock (HFR) method by \citet{pal03b} and \citet{men04}, differing on average by less than 0.1\%, and the dispersion of the radiative transition probabilities is not larger than 20\%. Regarding plasma effects, in Table~\ref{rad}, we tabulate the wavelengths and $A$-values for the stronger K lines ($A_{ki}\geq 10^{13}$\,s$^{-1}$) at $\mu = 0$, 0.1, and 0.25\,au, where it can be seen that they are hardly modified---redshifted by $\sim 1{-}2$\,m\AA\ or less---and the radiative rates only vary by a few percent in most cases (15--20\% in a handful of transitions).

\begin{figure}[!ht]
  \centering
  \includegraphics*[pagebox=mediabox, width=\columnwidth]{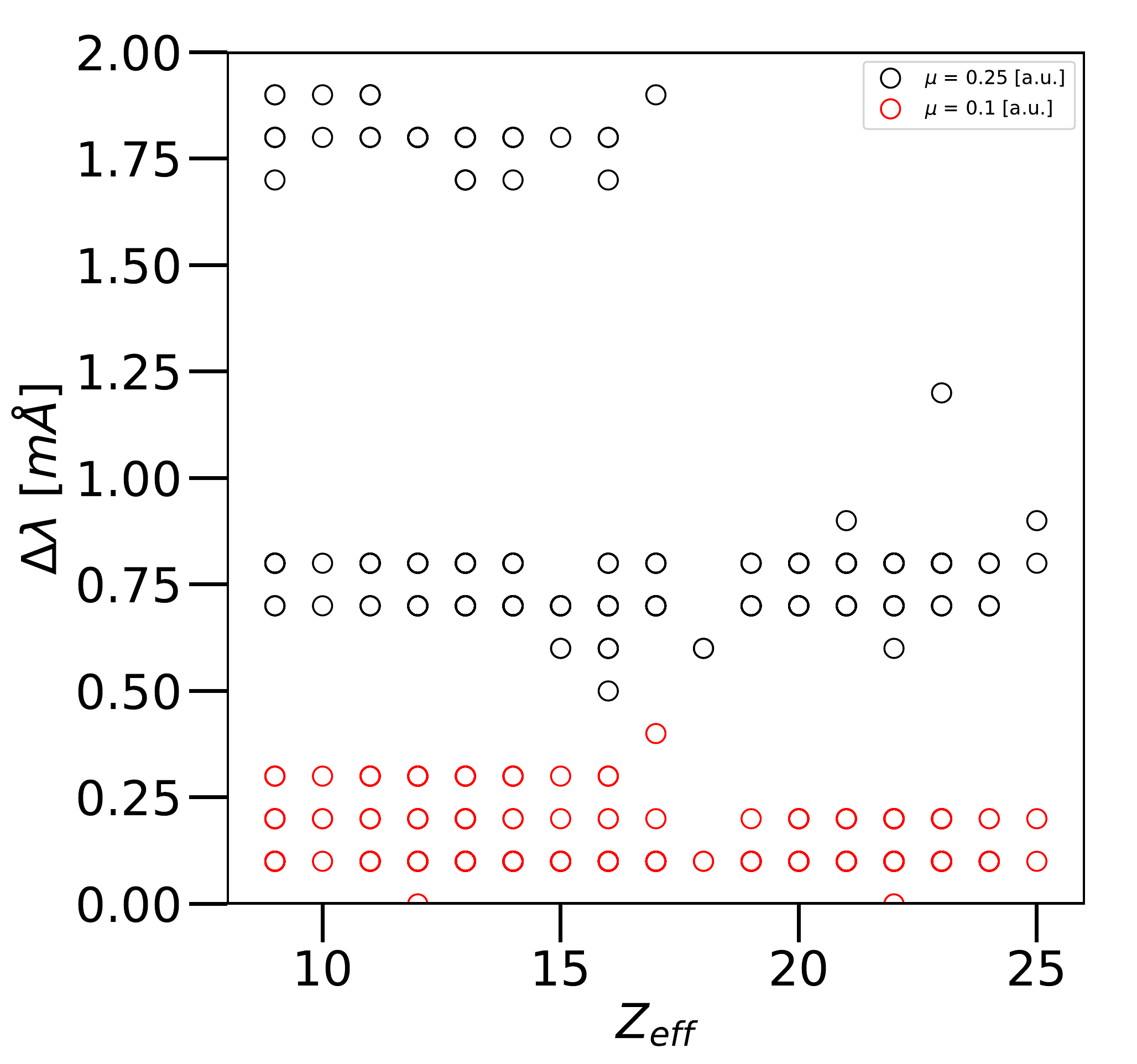}
  \caption{Wavelength shifts, $\Delta \lambda$, for K lines in \ion{Fe}{ix} -- \ion{Fe}{xxv} as a function of the effective charge $Z_{\rm eff}=Z-N+1$. Red circles: $\mu=0.1~{\rm au}$. Black circles: $\mu=0.25~{\rm au}$.} \label{Fig5}
\end{figure}

In Figure~\ref{Fig5}, we plot the wavelength shifts as a function of the ionic effective charge $9\leq Z_{\rm eff}\leq 25$ for $\mu=0.1$ and 0.25~au, again including the data from \citetalias{dep19b} for \ion{Fe}{xvii} -- \ion{Fe}{xxv}. We do not see a well-defined trend with $Z_{\rm eff}$, but for $Z_{\rm eff}\leq 17$, the K$\beta$ redshifts at $\mu=0.25$~au are found to be $\sim 2$\,m\AA; that is, a factor of 2 larger than the K$\alpha$ lines.

\subsection{Auger widths}

The Auger widths for the K-vacancy states we computed for the isolated atom are in good agreement with those obtained by \citet{pal03b} and \citet{men04}, differences being no large than 10--15\%. Similarly to the radiative rates, our Auger widths are only weakly modified by the plasma environment (see Table~\ref{auger}), with the reductions at $\mu  = 0.25$ au no greater than a few percent ($< 3\%$) with respect to the isolated atom.

\section{Summary and conclusions}

Following our previous work on the oxygen isonuclear sequence (Paper~I) and on the highly charged iron ions (Paper~II), we studied the influence of the plasma environment on the atomic structure and on the radiative and Auger parameters associated with the K lines of the moderately charged species \ion{Fe}{ix} -- \ion{Fe}{xvi}. The IPs, K-thresholds, wavelengths, radiative transition probabilities, and Auger widths have been calculated with the MCDF method, which includes a Debye-H\"uckel potential to model the plasma screening effects. To simulate the plasma conditions in compact-object accretion disks, we varied the plasma screening parameter $\mu$ from 0 au to 0.25 au.
Our main results are summarized as follows.
\begin{itemize}
    \item[1.] We confirmed the linear redshifts of both the IP and K threshold with the effective ionic charge and their close magnitude for each species. Their values range between $-20$~eV to $-100$~eV.
    \item[2.] The radiative and Auger rates are only marginally modified by the plasma environment (by a few percent overall).
    \item[3.] The transition wavelengths are redshifted by less than 1 m\AA, but for the most extreme conditions studied ($\mu= 0.25$~au), the more sensitive  K$\beta$ lines are shifted to the red by about $-2$~m\AA.
\end{itemize}

As discussed in the context of the highly ionized Fe ions (Paper~II), the high-density plasma effects presented here are expected to have a noticeable impact on the modeling of astrophysical environments where the temperature and density reach high enough values. A prime example is the material in accretion disks formed around compact objects such as white dwarfs, neutron stars, and black holes. The standard $\alpha$-disk model \citep{sha73} predicts densities in the mid-plane of the disk of order ${\sim}10^{13}{-}10^{27}$~cm$^{-3}$ for black holes with masses in the range $10{-}10^9$~M$_{\sun}$ and reasonable values of the accretion rate and viscosity. Recent simulations based on general relativity and magnetic hydrodynamics tend to agree with the orders of magnitude of these predictions, with lower mass objects having larger densities \citep[e.g.,][]{sch13,jia19a}. Accreting black holes are observed to typically emit large amounts of X-rays with a fraction of this flux reprocessed in the disk itself. This reprocessing produces a rich spectrum of fluorescence lines and other atomic features  commonly described as the X-ray reflection spectrum \citep{ros05,gar10}. The ubiquitousness of these signatures has made X-ray reflection spectroscopy a common and powerful technique for obtaining physical information  regarding the composition and ionization state of the accretion flow and the coveted black-hole spin \citep{fab89,rey19}. Due to their abundance and comparatively large fluorescence yield, along with their crucial role in determining the ionization structure of the gas producing most of the reprocessed signal, the iron ionic species featured in this paper are of particular interest for reflection modeling \citep{gar13}.

The high-quality observational data available in the last decade have revealed a number of important issues in the spectral modeling of accreting black holes; for instance, the modeling of the so-called ``soft excess'' in the spectrum active galactic nuclei (AGN). This is a broad and featureless component observed at soft X-ray energies ($\sim$1 keV), which can also be  described as such by several physically distinct models. As was recently demonstrated for the Seyfert~1 AGN Mrk 509 \citep{gar19}, one such model is based on the X-ray reflection spectrum from a very dense accretion disk. Higher disk densities lead to a strong excess of the reflected continuum at soft energies due to the enhancement of free--free heating in the atmosphere of the disk, which increases with rising density \citep{bal04,gar16}. An extended study of 17 AGN characterized by a strong soft excess have made use of high-density reflection models to reproduce their X-ray spectra \citep{jia19b}, finding tentative correlations between the black-hole mass, accretion rate, and accretion-disk density.

In the study of accreting sources, the nominal values of some of the parameters obtained in reflection modeling are also relevant: in particular, the iron abundance is commonly found to be larger than the expected Solar value by factors of two up to ten \citep{gar18}. This trend is found in accreting black holes with very different masses suggesting an artificial origin rather than a physical overabundance. Reflection calculations at densities higher than the traditionally assumed values (${\sim}10^{15}$~cm$^{-3}$) have shown promise in addressing both issues. Recent analyses of the reflection spectra from AGN and black-hole binaries (BHB) appear to indicate that the high-density effects are acting positively to resolve the mystery of the high iron abundances leading to substantially lower observed values \citep{tom18,jia19b,jia19c}.

The plasma environment effects presented here have not yet been included in the studies described above. However, they are expected to affect further the atomic features imprinted in the reflection spectra and, consequently, to modify the quantities deduced from the fits. Even though a comprehensive analysis of the propagation of plasma effects in synthetic spectra is outside the scope of the present work, our conclusions suggest that some of the spectral profiles observed in the X-ray reflected spectrum from accreting sources may be modified to an extent that could potentially influence their characterization. This will be relevant for the next generation of X-ray observatories such as {\it XRISM} \citep{tas18} and {\it ATHENA} \citep{nan13}, which will provide an enhanced effective area in the X-ray band and instrumentation with far superior spectral resolution (of the order of a few eV) to aid in the detection of the lowering of the ionic K edges in extreme plasma conditions.

%


\begin{acknowledgements}
JD is Research Fellow of the Belgian Fund for Research Training in Industry and Agriculture (FRIA) while PP and PQ are, respectively, Research Associate and Research Director of the Belgian Fund for Scientific Research (F.R.S.-FNRS). Financial supports from these organizations, as well as from the NASA Astrophysics Research and Analysis Program (grant 80NSSC17K0345) are gratefully acknowledged. JAG acknowledges support from the Alexander von Humboldt Foundation.
\end{acknowledgements}

\begin{table*}[!ht]
  \caption{Wavelengths and transition probabilities for the K lines of \ion{Fe}{ix} -- \ion{Fe}{xvi} ($9\leq Z_{\rm eff}\leq 16$) computed with plasma screening parameters $\mu = 0.0$, 0.1, and 0.25~au. Data obtained with $\mu = 0$~au correspond to the isolated atom. \label{rad}}
  \centering
  \tiny
  \begin{tabular}{clccccccc}
  \hline\hline
  \noalign{\smallskip}
  $Z_{\rm eff}$ & Transition & \multicolumn{3}{c}{Wavelength (\AA)} && \multicolumn{3}{c}{Transition probability (s$^{-1}$)} \\
  \cline{3-5} \cline{7-9}
  \noalign{\smallskip}
     &  & $\mu = 0.0$ & $\mu = 0.1$ & $\mu = 0.25$ & & $\mu = 0.0$ & $\mu = 0.1$ & $\mu = 0.25$ \\
  \hline
  \noalign{\smallskip}
  9      & $\mathrm{[1s]3d~^3D_2 - [3p]3d~^3F_3}$             & 1.7538 & 1.7541 & 1.7556 & & 4.143E+13 & 4.146E+13 & 4.164E+13 \\
  9      & $\mathrm{[1s]3d~^3D_1 - [3p]3d~^3F_2}$             & 1.7540 & 1.7542 & 1.7558 & & 5.446E+13 & 5.450E+13 & 5.472E+13 \\
  9      & $\mathrm{[1s]3d~^1D_2 - [3p]3d~^3D_3}$             & 1.7546 & 1.7549  & 1.7565 & & 1.961E+13 & 1.958E+13 & 1.943E+13 \\
  9      & $\mathrm{[1s]3d~^3D_2 - [3p]3d~^1D_2}$             & 1.7548 & 1.7550  & 1.7566 & & 2.961E+13 & 2.961E+13 & 2.961E+13 \\
  9      & $\mathrm{[1s]3d~^1D_2 - [3p]3d~^3D_2}$             & 1.7548 & 1.7551  & 1.7567 & & 2.518E+13 & 2.518E+13 & 2.527E+13 \\
  9      & $\mathrm{[1s]3d~^3D_1 - [3p]3d~^3D_1}$             & 1.7549 & 1.7551  & 1.7567 & & 2.460E+13 & 2.464E+13 & 2.480E+13 \\
  9      & $\mathrm{[1s]3d~^1D_2 - [3p]3d~^1F_3}$             & 1.7549 & 1.7552  & 1.7567 & & 2.334E+13 & 2.342E+13 & 2.381E+13 \\
  9      & $\mathrm{[1s]3d~^1D_2 - [3p]3d~^1P_1}$             & 1.7588 & 1.7590  & 1.7605 & & 1.556E+13 & 1.559E+13 & 1.572E+13 \\
  9      & $\mathrm{[1s]3d~^3D_1 - [2p]3d~^3P_0}$             & 1.9357 & 1.9358  & 1.9365 & & 6.334E+13 & 6.332E+13 & 6.322E+13 \\
  9      & $\mathrm{[1s]3d~^1D_2 - [2p]3d~^3P_1}$             & 1.9358 & 1.9359  & 1.9366 & & 1.997E+13 & 1.993E+13 & 1.973E+13 \\
  9      & $\mathrm{[1s]3d~^3D_1 - [2p]3d~^3P_1}$             & 1.9359 & 1.9360  & 1.9367 & & 1.061E+14 & 1.060E+14 & 1.057E+14 \\
  9      & $\mathrm{[1s]3d~^3D_2 - [2p]3d~^3P_1}$             & 1.9359 & 1.9360  & 1.9367 & & 3.040E+13 & 3.044E+13 & 3.064E+13 \\
  9      & $\mathrm{[1s]3d~^1D_2 - [2p]3d~^3P_2}$             & 1.9362 & 1.9363  & 1.9370 & & 8.137E+13 & 8.131E+13 & 8.098E+13 \\
  9      & $\mathrm{[1s]3d~^3D_1 - [2p]3d~^3P_2}$             & 1.9363 & 1.9364  & 1.9371 & & 2.921E+13 & 2.912E+13 & 2.868E+13 \\
  9      & $\mathrm{[1s]3d~^3D_2 - [2p]3d~^3P_2}$             & 1.9363 & 1.9364  & 1.9371 & & 3.217E+13 & 3.221E+13 & 3.239E+13 \\
  9      & $\mathrm{[1s]3d~^3D_2 - [2p]3d~^3F_3}$             & 1.9365 & 1.9366  & 1.9372 & & 2.690E+14 & 2.689E+14 & 2.684E+13 \\
  9      & $\mathrm{[1s]3d~^1D_2 - [2p]3d~^1D_2}$             & 1.9367 & 1.9368  & 1.9375 & & 4.062E+13 & 4.062E+13 & 4.064E+13 \\
  9      & $\mathrm{[1s]3d~^3D_1 - [2p]3d~^1D_2}$             & 1.9368 & 1.9370  & 1.9376 & & 1.554E+14 & 1.554E+14 & 1.555E+14 \\
  9      & $\mathrm{[1s]3d~^3D_2 - [2p]3d~^1D_2}$             & 1.9368 & 1.9369  & 1.9376 & & 4.873E+13 & 4.867E+13 & 4.837E+13 \\
  9      & $\mathrm{[1s]3d~^1D_2 - [2p]3d~^3D_3}$             & 1.9369 & 1.9370  & 1.9377 & & 1.297E+14 & 1.297E+14 & 1.294E+14 \\
  9      & $\mathrm{[1s]3d~^1D_2 - [2p]3d~^3D_1}$             & 1.9381 & 1.9383  & 1.9389 & & 6.805E+13 & 6.799E+13 & 6.765E+13 \\
  9      & $\mathrm{[1s]3d~^3D_1 - [2p]3d~^3D_1}$             & 1.9383 & 1.9384  & 1.9390 & & 6.597E+13 & 6.597E+13 & 6.598E+13 \\
  9      & $\mathrm{[1s]3d~^3D_1 - [2p]3d~^3F_2}$             & 1.9402 & 1.9403  & 1.9409 & & 1.280E+14 & 1.280E+14 & 1.275E+14 \\
  9      & $\mathrm{[1s]3d~^3D_2 - [2p]3d~^3F_2}$             & 1.9402 & 1.9403  & 1.9409 & & 1.078E+14 & 1.077E+14 & 1.075E+14 \\
  9      & $\mathrm{[1s]3d~^1D_2 - [2p]3d~^3D_2}$             & 1.9402 & 1.9403  & 1.9410 & & 6.681E+13 & 6.680E+13 & 6.675E+13 \\
  9      & $\mathrm{[1s]3d~^1D_2 - [2p]3d~^3F_3}$             & 1.9403 & 1.9404  & 1.9411 & & 1.367E+14 & 1.366E+14 & 1.364E+14 \\
  9      & $\mathrm{[1s]3d~^1D_2 - [2p]3d~^1P_1}$             & 1.9419 & 1.9420  & 1.9427 & & 2.439E+13 & 2.446E+13 & 2.479E+13 \\
  9      & $\mathrm{[1s]3d~^3D_1 - [2p]3d~^1P_1}$             & 1.9420 & 1.9421  & 1.9428 & & 1.776E+13 & 1.776E+13 & 1.775E+13 \\
  9      & $\mathrm{[1s]3d~^3D_2 - [2p]3d~^1P_1}$             & 1.9420 & 1.9421  & 1.9428 & & 7.565E+13 & 7.555E+13 & 7.503E+13 \\
  10 & $\mathrm{[1s]3p^6~^2S_{1/2} - 3p^5~^2P_{3/2}}$     & 1.7519 & 1.7521 & 1.7537 & & 7.603E+13 & 7.610E+13 & 7.647E+13 \\
  10 & $\mathrm{[1s]3p^6~^2S_{1/2} - 3p^5~^2P_{1/2}}$     &     1.7523 & 1.7526 & 1.7542 & & 3.677E+13 & 3.681E+13 & 3.698E+13 \\
  10 & $\mathrm{[1s]3p^6~^2S_{1/2} - [2p]3p^6~^2P_{3/2}}$ &     1.9367 & 1.9369  & 1.9375 & & 3.830E+14 & 3.828E+14 & 3.822E+14 \\
  10 & $\mathrm{[1s]3p^6~^2S_{1/2} - [2p]3p^6~^2P_{1/2}}$ &     1.9405 & 1.9406  & 1.9412 & & 1.880E+14 & 1.880E+14 & 1.877E+14 \\
  \hline
  \end{tabular}
  \tablefoot{A complete version of this table is available in electronic form from the CDS.}
\end{table*}

\begin{table*}[!ht]
  \caption{Plasma environment effects on the Auger widths of K-vacancy states in \ion{Fe}{ix} -- \ion{Fe}{xvi}  ($9\leq Z_{\rm eff}\leq 16$) computed with plasma screening parameters $\mu = 0.0$, 0.1, and 0.25~au. Data obtained with $\mu = 0$~au. correspond to the isolated atom. \label{auger}}
  \centering
  \small
  \begin{tabular}{clccc}
  \hline\hline
  \noalign{\smallskip}
  $Z_{\rm eff}$ & Level & \multicolumn{3}{c}{Auger width (s$^{-1}$)} \\
  \cline{3-5}
  \noalign{\smallskip}
    &  &  $\mu = 0.0$ & $\mu = 0.1$  & $\mu = 0.25$ \\
  \hline
  \noalign{\smallskip}  
9       &       $\mathrm{[1s]3d~^3D_1}$         &       1.057E+15       &       1.054E+15       &       1.051E+15       \\
9       &       $\mathrm{[1s]3d~^3D_2}$         &       1.056E+15       &       1.053E+15       &       1.050E+15       \\
9       &       $\mathrm{[1s]3d~^3D_3}$         &       1.056E+15       &       1.053E+15       &       1.050E+15       \\
9       &       $\mathrm{[1s]3d~^1D_2}$         &       1.056E+15       &       1.053E+15       &       1.050E+15       \\
10      &       $\mathrm{[1s]3p^6~^2S_{1/2}}$   &       9.832E+14       &       9.789E+14       &       9.747E+14       \\
11      &       $\mathrm{[1s]3p^5~^3P_2}$       &       1.252E+15       &       1.244E+15       &       1.230E+15       \\
11      &       $\mathrm{[1s]3p^5~^3P_1}$       &       1.251E+15       &       1.243E+15       &       1.230E+15       \\
11      &       $\mathrm{[1s]3p^5~^3P_0}$       &       1.254E+15       &       1.246E+15       &       1.232E+15       \\
11      &       $\mathrm{[1s]3p^5~^1P_1}$       &       1.235E+15       &       1.227E+15       &       1.213E+15       \\
12      &       $\mathrm{[1s]3p^4~^4P_{5/2}}$   &       1.257E+15       &       1.254E+15       &       1.247E+15       \\
12      &       $\mathrm{[1s]3p^4~^4P_{3/2}}$   &       1.256E+15       &       1.254E+15       &       1.246E+15       \\
12      &       $\mathrm{[1s]3p^4~^4P_{1/2}}$   &       1.259E+15       &       1.256E+15       &       1.248E+15       \\
12      &       $\mathrm{[1s]3p^4~^2P_{3/2}}$   &       1.231E+15       &       1.228E+15       &       1.220E+15       \\
12      &       $\mathrm{[1s]3p^4~^2P_{1/2}}$   &       1.226E+15       &       1.224E+15       &       1.216E+15       \\
12      &       $\mathrm{[1s]3p^4~^2D_{5/2}}$   &       1.248E+15       &       1.246E+15       &       1.238E+15       \\
12      &       $\mathrm{[1s]3p^4~^2D_{3/2}}$   &       1.247E+15       &       1.244E+15       &       1.237E+15       \\
12      &       $\mathrm{[1s]3p^4~^2S_{1/2}}$   &       1.247E+15       &       1.244E+15       &       1.237E+15       \\
  \hline        
  \end{tabular}
  \tablefoot{A complete version of this table is available in electronic form from the CDS.}
\end{table*}

\end{document}